\documentclass[12pt,preprint]{aastex}
\usepackage{graphicx}
\usepackage{amssymb}
\begin{document}

\title{Possible Optical Detection of the Anomalous X-ray Pulsar
CXOU~J010043.1$-$721134}
\author{Martin Durant and Marten H. van Kerkwijk}
\affil{Department of Astronomy and Astrophysics, University of
  Toronto\\  60 St. George St, Toronto, ON\\ M5S 3H8, Canada }

\begin{abstract}
Archival Hubble Space Telescope Wide Field/Planetary Camera 2
observations of the Small Magellanic Cloud serendipitously
reveal a possible counterpart to the Anomalous X-ray Pulsar
CXOU~J010043.1$-$721134. The candidate 
is faint, but its location and strange colours make it an
interesting object. We estimate, that the 
probability of such a detection being due to a non-physical
source is less than 1.5\%. 
We have tried to confirm the identification with Gemini-South and Magellan,
but the conditions were insufficiently favourable.
If confirmed, the object will allow the first detailed studies of the
optical and ultraviolet emission of magnetars.
\end{abstract}
\maketitle

\section{Introduction}
The anomalous X-ray pulsars (AXPs) are a class of neutron stars,
numbering about half a dozen, which are radio-quiet, with periods of
the order $\sim10$s and estimated ages of $10^3$ to $10^5$yr. Like the soft
gamma-ray repeaters, they
are thought to be {\em magnetars}, whose emission is powered by the decay
of a super-strong magnetic field ($\sim 10^{15}$G). See Woods \&
Thompson (2004) for a review of the known magnetars and their
properties. 

While energetically, the emission at X-ray energies dominates, optical
and infrared photometry of AXPs is giving interesting constraints on the
physical processes of the stellar magnetospheres. Particularly
intriguing is that for the brightest object, 4U~0142+61, the optical
spectral energy distribution is not just a power law. It shows, unique
among neutron stars, a spectral break between V and B (Hulleman et
al., 2004).  Unfortunately, because of the
uncertainty in the high amount of reddening, the precise shape cannot
be measured. 

In the
magnetar model, the optical emission could be dues to ion cyclotron
emission. If so, the spectral break should be a general feature
(C. Thompson, 2004, priv comm.) due to the existence of a {\em cooling
  radius} in the magnetar magnetosphere from within which ions do not
radiate (for a brief discussion, see Hulleman et al., 2004).
The $\sim5$ other AXPs known so far are, unfortunately, too highly
reddened to be detected
in V or B. Another prediction is that the spectra of different AXPs should be
similar, but again uncertainties in the reddening do not allow us
to test this (e.g. Durant \& van Kerkwijk, 2005). As an alternative
model, Eichler et al. (2002), considered the possibility of coherent
optical and infrared emission from the lower magnetosphere of a
magnetar, in analogy to some radio pulsar
models. Unfortunately, no clear predictions for the spectral shape
were made.

For the purposes of investigating the optical spectra of AXPs, the
recent discovery of an AXP in the Small Magellanic Cloud (SMC),
CXOU~J010043.1$-$721134 (Lamb et al., 2002; Majid
et al., 2004) is particularly interesting.
It is the only AXP found so far, that is not confined to the disc of the Milky
Way. The reddening to this source is, therefore, much less than for
the other AXPs. Furthermore, its distance is relatively well known at 60.6(1.0)
kpc (e.g. Hilditch et al., 2005). It thus presents a unique opportunity to study
an AXP in the blue/UV.

\section{Archival Observation and Analysis}

Seeking imaging data on CXOU~J010043.1$-$721134, we searched all the
archives available t us. We found that the field
was observed on 20 April 2004 with
the Wide Field and Planetary Camera 2 (WFPC2) on board the Hubble
Space Telescope (HST), as part of a snapshot programme for
three-colour photometry of several patches of the SMC (Tolstoy, 1999).
Single exposures were taken of 230s in the near-ultraviolet F300W,
180s in the ``broad V'' F606W and 300s in Cousins I-like F814W filters.
The position of our object of interest is on
chip WF2 of the WFPC2 array.

We determined an astrometric solution by matching sources off the WF2
image to objects in the USNO B1.0 catalogue (Monet et al., 2003), and
fitting for offset, rotation and scale. Eight stars were matched,
after rejecting 7 objects which had poorly measured positions or which
corresponded to multiple sources on the WF2 image. With these eight
sources, the uncertainty in the astrometric fit is $0\farcs19/\sqrt6 =
0\farcs08$ in each co-ordinate for the F606W frame. The uncertainty in
applying the astrometry to the other two bands was negligible in
comparison. The systematic uncertainty in connecting the USNO
astrometry to the International Celestial Reference System is
$0\farcs2$ in each co-ordinate, and the
uncertainty in the {\em Chandra} position of CXOU~J010043.1$-$721134
is a radius $r=0\farcs6$ at 90\% confidence. Note that the latter is
from the nominal {\em Chandra} performance, despite being somewhat
off-axis (Lamb et al., 2002). The above numbers, combined in
quadrature, give a total uncertainty in the AXP's position on our images of
$r=0\farcs72$ at 90\% confidence.  Photometry was performed using {\tt
HSTphot 1.1} (Dolphin, 2000).

\begin{figure}
\begin{center}
\includegraphics[width=0.75\hsize,angle=270]{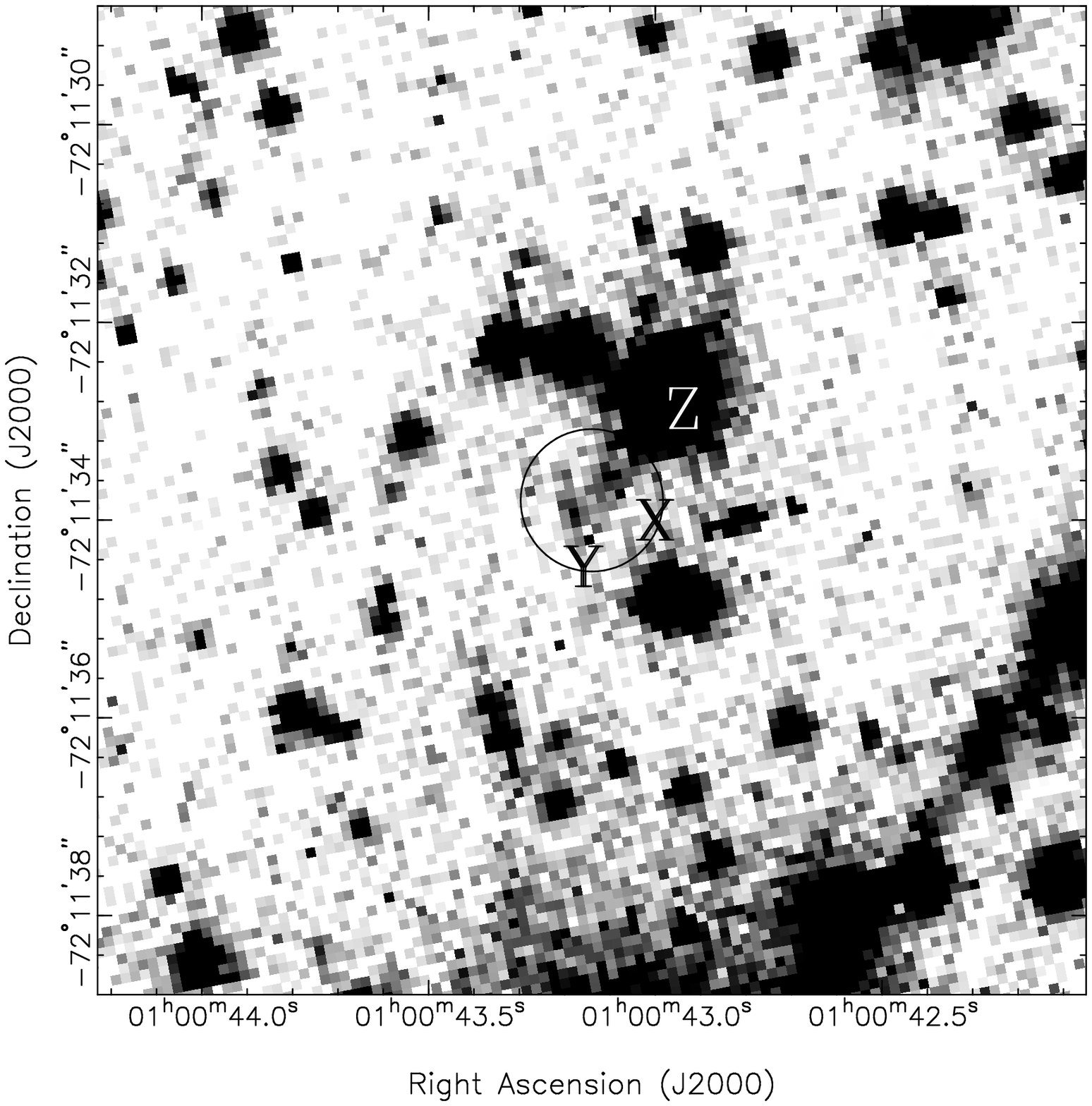}
\parbox{0.45\hsize}{\includegraphics[width=\hsize,angle=270,clip=]{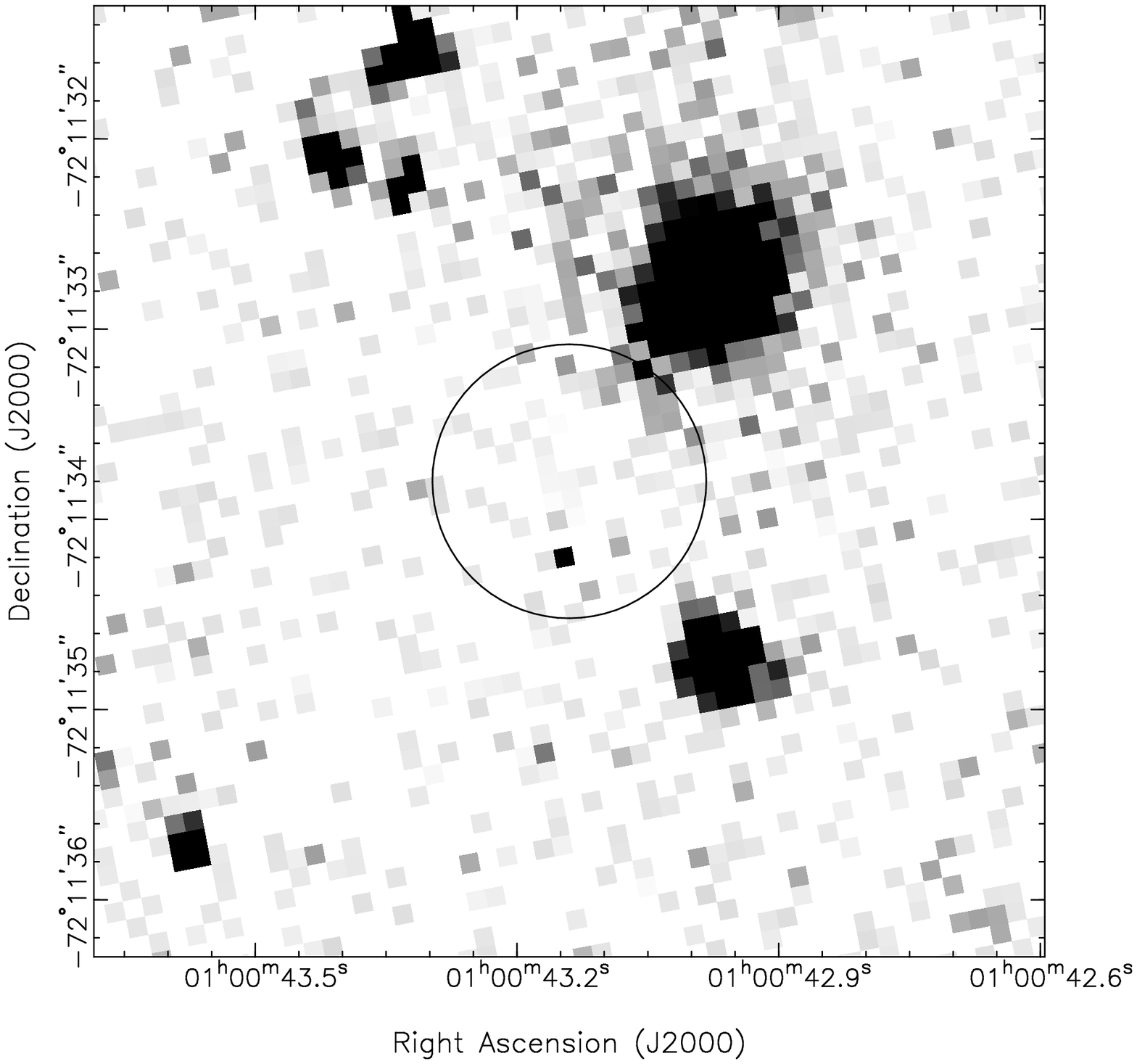}}
\parbox{0.45\hsize}{\includegraphics[width=\hsize,angle=270,clip=]{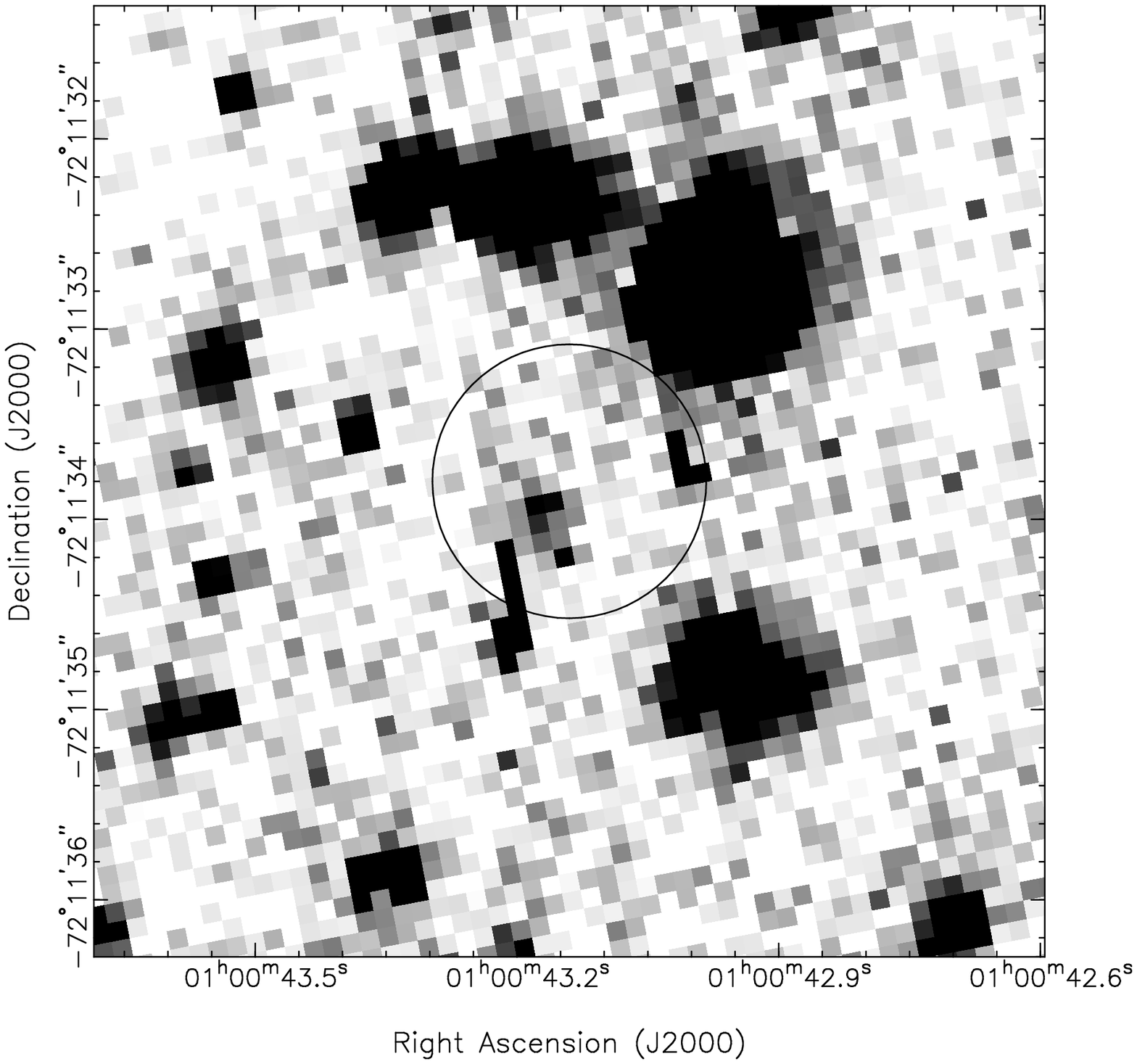}}
\caption{WFPC2 images of the field of CXOU~J010043.1$-$721134, in F606W
  (top), F300W (bottom left) and F814W (bottom right). Star X is
  the proposed counterpart, and Stars Y and Z its nearest
  neighbours. The circle shows the uncertainty in the Chandra position at
  90\% confidence.}\label{field}
\end{center}
\end{figure}

\begin{figure}
\includegraphics[width=\hsize]{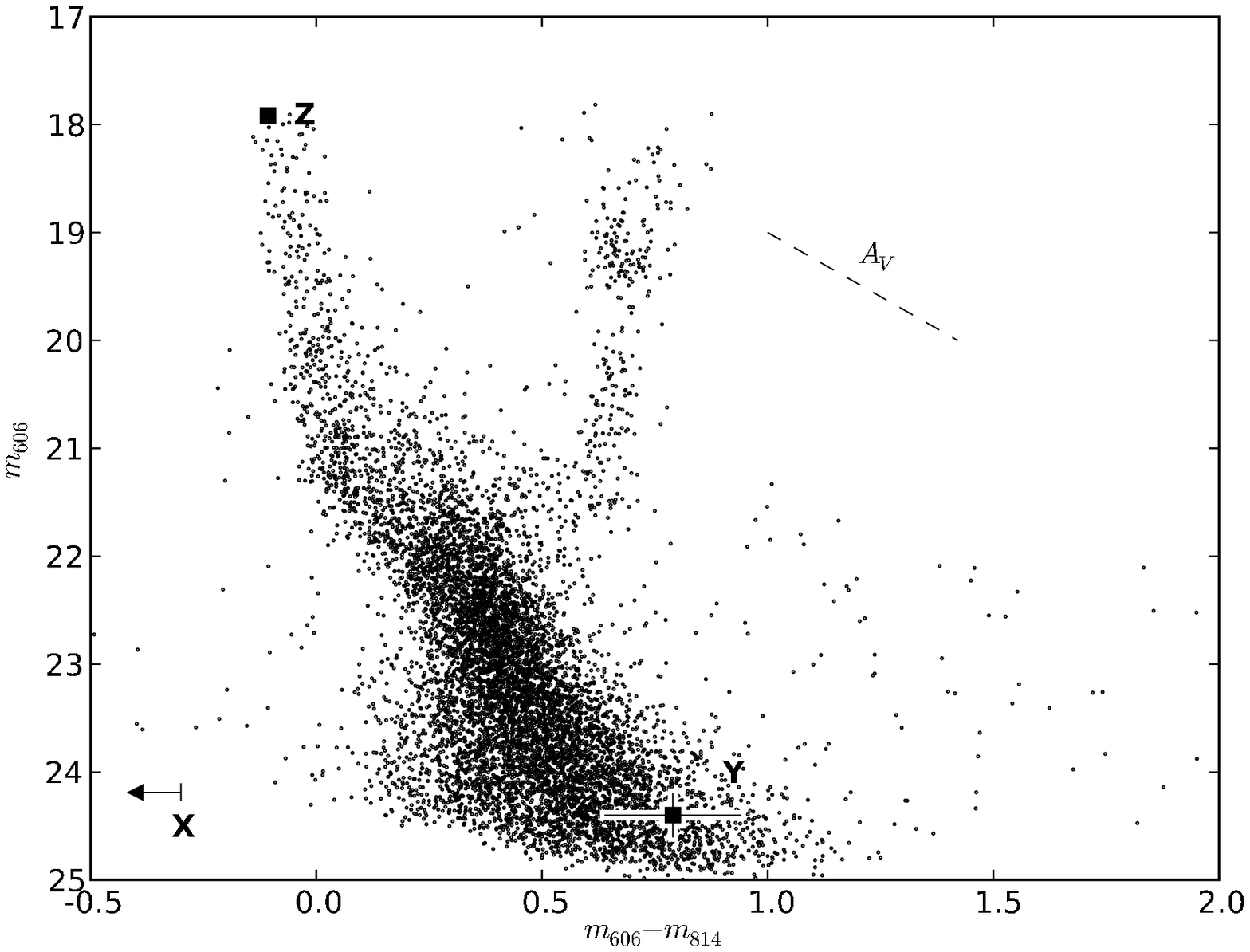}
\caption{Colour-magnitude diagramme of the field around CXOU
  J010043.1$-$721134. Stars Y and Z are labelled, and Star X is shown as
  a limit. The effect of one magnitude of visual reddening is shown by
  the dashed line.}\label{CMD}
\end{figure}

Figure \ref{field} shows the F606W image of the field immediately
around CXOU~J010043.1$-$721134, with the positional error circle
indicated. Stars X and Y have positions
consistent with that of the AXP, with Star Z being a nearby, much
brighter source. Their positions and magnitudes are listed in Table
\ref{phot}, and indicated in a
colour-magnitude diagramme of all stars detected in the WFPC2 images in Figure
\ref{CMD}. 

From the photometry, Star Y is consistent with being a G5V star
at the distance and reddening of the SMC, and Star Z an early
B-type star. The colours and magnitudes of Star X do not
correspond to any known stellar type, and make it a clear out-lier in Figure
\ref{CMD}, suggesting a very blue, possibly hot object. Based on its
position and
unusual colours, we therefore consider Star X a likely counterpart to
CXOU~J010043.1$-$721134.

\begin{deluxetable}{lccccccc}
\tablecaption{Astrometry and photometry of stars near CXOU
  J010043.1$-$721134 \label{phot}}
\tablewidth{0pt}
\tablehead{ \colhead{Star} & \colhead{R.A.$_{J2000}$} & \colhead{Dec$_{J2000}$} &
  \colhead{$m_{300}$} & \colhead{$m_{606}$}
  & \colhead{$m_{814}$} & \colhead{$m_{606}-m_{814}$} &
  \colhead{$M_V$\tablenotemark{a}}}
\startdata
X\tablenotemark{b} & 01:00:43.109 & -72:11:33.77 & $>21.7$ & 24.19(15) & $>24.5$ & $<-0.3$ & 5.0 \\
Y & 01:00:43.187 & -72:11:34.14 & $>21.7$ & 24.40(15) & 23.61(14) & 0.8(2) & 5.2 \\
Z & 01:00:42.990 & -72:11:33.01 & 16.295(8) & 17.915(4) & 18.022(7) & -0.107(8) & -2.9 \\
\enddata
\tablecomments{~Limits are at the $3\sigma$ level.}
\tablenotetext{a}{Calculated using $(m-M)_0=18.9$ and $A_V=0.3$ (Hilditch et
  al., 2004) and assuming $m_{606}\simeq V$}
\tablenotetext{b}{Proposed counterpart to CXOU~J010043.1$-$721134,
  which has position $R.A.=01:00:43.14$, $dec=-72:11:33.8$}
\end{deluxetable}

As a caveat, however , it should be remembered that this measurement is based on
a single F606W exposure. The source in Figure \ref{field} does not
appear like a cosmic ray hit, and the {\tt HSTphot} $\chi$ and $Sharp$
parameters are within reasonable limits for a point source: $\chi=1.18$
(goodness of fit parameter; reasonable values: $<2.5$) and
$Sharp=-0.425$ (where 0 corresponds to a stellar point spread
function, positive values to more peaked profiles and negative values
to more diffuse ones; reasonable values:
-0.5-- 0.5). There were no bad pixels within Star X's
profile. 

In order to test the robustness of this identification, we estimated
the likelihood of such a detection in an error circle of this size at
any point on this chip of the detector.
We searched the photometry for all objects in the F606W
image which are classified as stellar ($\chi<2.5$, $-0.5<sharp<0.5$), and
with $m_{606}-m_{814}<-0.3$, i.e. at least as blue as Star X.
Forty such objects are found on the same chip as
Star X (including those near Star X in Figure \ref{CMD}), giving the
probability of one falling within a circle of 
radius $0\farcs72$ of $\approx1.5$\%. We note that the majority of
these are within 3 pixels of brighter sources in the F814W
image, and consequently were not detected in that band. Since this does not
apply to Star X, and there do not appear to be any artifacts close to
it (i.e. the location of Star X appears like
sky in the F814W image), the chance of it being a false detection is somewhat smaller,
but how much smaller is hard to quantify. 

We also calculated the likelihood of our putative detection being due to a
cosmic ray hit or instrumental effects. We searched for objects
which are classified as stars in terms of their
{\em Chi} and {\em Sharp} parameters as above, which were detected in
F300W but not in F606W. We find thirteen objects, which implies that the
probability of Star X being due to a cosmic rays hit or
purely instrumental effects is 0.4\% (after correcting for the difference in
exposure time between F300W and F606W). 

Seeking to confirm this detection, we searched other archives for
optical images. We found a V-band image from the Wide Field Imager
(WFI) on the 2.2m ESO telescope, La Silla, Chile. This demonstrated
that the area of sky
was very crowded, and extremely good seeing would be required to
separate and securely detect Star X. In this case, the seeing was
poor. We also obtained Gemini DDT observations 
with GMOS-S (Crampton \& Murowinski, 2004) at Cerro Pachon,
Chile. Unfortunately, the seeing was also not good enough in these
images to distinguish between the sources in the crowded field. A
proposal was also accepted at Magellan, Las Campanas, Chile, but
conditions have not been good enough to obtain images so far.
Unfortunately, the presence of Star Z means that only the most
exceptional seeing conditions will allow further measurements of this
object from the ground.

\section{Discussion and Conclusions}
Taking Star X as the true optical counterpart, CXOU~J010043.1$-$721134
has an X-ray to optical flux ratio
$F_X/F_V=1.0\times10^{-13}/5.5\times10^{-15}=18$ (un-absorbed X-ray flux in the
2--10keV range from Woods \& Thompson, 2004; visual $\nu F_\nu$ flux is
de-reddened using $A_V=0.3$ [Hilditch et al., 2005], and assumes
$m_{606}=V$). This compares with $F_X/F_V=460$ for 4U~0142$+$61 (for
$A_V=5.1$, the nominal reddening), the only
other AXP with an optical detection (Hulleman et al., 2004).
Clearly the two ratios are very different.

It has been observed that infrared to X-ray flux ratios are similar
for those AXPs with secure measurements (4U~0142$+$61,
1E~1048.1$-$5937 and 1E~2259$+$586; Durant
\& van Kerkwijk, 2005). Variations have, however, been observed to be
very large, of orders of magnitude in some cases. For example the {\em
  transient AXP} XTE~J1810-197 (Ibrahim et
al, 2004) increased dramatically in both X-rays and infrared flux
before slowly dimming again. 

It is possible that the
difference in V-band to X-ray flux ratio above arises because the
measurements for CXOU~J010043.1$-$721134
were not simultaneous; the AXP could have been brighter by a large
factor at the time of the HST observation. 
CXOU~J010043.1$-$721134 was observed to be $sim50$\% brighter in
X-rays by Majid et al. (2004) than Lamb et al. (2002), but they
attribute this to the different instruments used to make the
observations rather than genuine variability.
4U~0142$+$61 has been the most stable of the AXPs in both X-ray and
optical flux (Hulleman et al., 2004).
This could, in principle, mean that the intrinsic spectra
of the two objects are very different, possibly indicating differing
magnetic field configurations.

The limit in F814W already provides some constraints on the shape of
the optical spectrum. Whilst a Rayleigh-Jeans form $\nu F_\nu
\sim \nu^n$, $n=3$ is possible, a flat spectrum ($n=0$) is
excluded. The 90\% confidence limit is $n\geq2$. Since the spectrum
should not increase steeper than
Rayleigh-Jeans (in the absence of an emission feature), we predict
that the I-band magnitude is not much below the limit we have
established. The F300W limit is not constraining in this respect.

In summary, we present Star X, with $m_{606}=24.19(15)$, as the probable optical
counterpart to CXOU~J010043.1$-$721134. It is at the right location and
has colours unlike normal stellar sources. Although based on a
detection in a single exposure, {\tt HSTphot} diagnostics point to it
being a real detection, with only a $\lesssim1.5$\% probability of a
false detection. If confirmed, this discovery will enable the measurement of AXP
properties in the blue and UV. 

\medskip\noindent{\bf Acknowledgements:}
This work made use of archival observations made with the NASA/ESA
Hubble Space Telescope and with observations from ESO Telescopes at the La
Silla Observatories.
We thank Slavek Rucinski and the Gemini Observatories for
attempting follow-up observations. We thank an anonymous referee for
very useful comments which much improved the presentation of our
results. We acknowledge financial support from NSERC.

\end{document}